\documentclass[12pt,amsmath,amsfonts]{JHEP3}

\title{The Low Energy Effective Equations of Motion for Multibrane Worlds Gravity \\}
\author{
Freddy P. Zen,$^a$ Arianto,$^{a,b}$
 Bobby E. Gunara,$^a$ Hishamuddin Zainuddin $^c$ \\

$^a$Theoretical Physics Laboratory, Department of Physics,\\
Institut Teknologi Bandung \\
Jl. Ganesha 10 Bandung 40132, Indonesia.\\

$^b$Department of Physics, Universitas Udayana  \\
Jl. Kampus Bukit Jimbaran Denpasar 80361, Indonesia. \\

$^c$Theoretical Studies Laboratory, ITMA, Universiti Putra Malaysia,\\
43400 UPM Serdang, Selangor, Malaysia. \\

E-mail: fpzen@fi.itb.ac.id, arianto@upk.fi.itb.ac.id,
bobby@fi.itb.ac.id, hisham@fsas.upm.edu.my. }

\preprint{}

\abstract{The three 3-brane system with both positive or negative
tension is studied in a low energy regime by using gradient
expansion method. The effective equations of motion on the brane
is derived and in particular we examine, in the first order, the
radion effective lagrangian for this system. In this case, we show
the solution of the modified Friedmann equation with dark
radiation on the middle brane and the other 3-branes by direct
elimination of the radion fields and Weyl scaling of the metric on
the branes. We also derived the scalar-tensor gravity on the
branes.}


\begin{document}

\section{Introduction}
One of the long standing problem in particle physics and
gravitational theories is how to understand quantum theory of
gravity. Nowadays, the only possible candidate for this theory is
the superstring theory~\cite{GSW}. Interestingly, this theory
predicts the existence of the extra dimensions. In order to
reconcile this prediction with our observed four dimensional
universe, we need a mechanism to compactify the extra dimensions.
In particular the setup of heterotic M-theory and its
compactification down to five dimensions~\cite{LOSW} leads to a
well motivated five dimensional brane world scenario, which can be
used to study its consequences in particle physics and cosmology.

Randall and Sundrum (RS) proposed two similar but distinct
phenomenological brane world scenarios~\cite{RS1, RS2}. The first
scenario is composed of two branes of the opposite tension, namely
RS I. This scenario is the five dimensional space-time which all
matter fields are assumed to be confined on branes at fixed points
of the $S^1/Z_2$ orbifold so that the bulk is described by pure
Einstein gravity with a negative cosmological constant. On the
other hand, the second scenario has a single brane with a positive
tension, namely RS II. The fifth dimension is infinite but still
$Z_2$ symmetry is imposed. In both scenarios the existence of the
branes and the bulk cosmological constant makes the bulk geometry
curved, or warped.

Furthermore, the brane world models are expected to shed some
light onto not only quantum gravity and unification issues, but
also cosmological issues such as the cosmological constant or dark
energy problem, or even particle physics ones such as the
hierarchy problem. There is also a brane world alternative to the
standard big bang plus inflation scenario~\cite{KOST, TTS}. This
scenario consist of a five dimensional bulk bounded by two branes,
which are as usual located at the fixed point of the $S^1/Z_2$
orbifold. Although many brane world models might be over
simplified, they should help in learning and understanding the
properties of an effective theory derived from some action in
space-time dimensions $d > 4$~\cite{FBAH}.

There are two approaches to obtain the effective Einstein
equations on the brane in the context of Randall-Sundrum
scenarios, namely covariant curvature formulation and gradient
expansion method. In the covariant formulation~\cite{SMS}, the
effective Einstein equation can be obtained by projecting the
covariant five dimensional Einstein equations on to the brane. The
resulting projected equations are modified with respect to general
relativity due to the presence of a local quadratic term in the
sources and to the presence of a non-local term which is the
projection of the five dimensional Weyl tensors. This last term
carries information of the bulk gravitational field on the brane
and its contribution is of fundamental importance as it might be
relevant even at low energy~\cite{Maartens1, Maartens2, Brax,
Langlois}.

The main difficulty in understanding the contribution of the
projected Weyl tensor to the effective theory on the brane is in
its non-local character. The equation for the projected Weyl
tensor on the brane are not closed so that solving the full five
dimensional equation of motion is necessary.

On the other hand, gradient expansion method gives a way out of
this problem. This method first proposed by Kanno and
Soda~\cite{KS1}. The main idea is to treat the issue
perturbatively, defining a low energy regime in which the energy
density on the brane is kept small with respect to bulk vacuum
energy density. The perturbation parameter is defined as the ratio
between these two energy densities. The five dimensional equations
of motion can be solved at different orders in the perturbation
parameter. This method allows in principle to derive the effective
equations of motion on the brane at each order.

In this paper, we generalize the results obtained for a two
3-brane scenario (RS I) to a multibrane scenario~\cite{LCotta, KK,
KSW} with the set up of the following. We consider the A, B, C
3-branes system, with A and B branes are placed at the fixed
points of the orbifold whereas C brane is put between the two
fixed points. We derive the effective equations of motion in this
scenario. By performing a perturbative expansion of the metric, an
expansion of the extrinsic curvature tensor and Weyl tensor are
considered. The four basic equations for the five dimensional
evolution equations and junction conditions are then solved at
different order in the expansion parameter. The parameter of
expansion is determined as in the Anti-deSitter (AdS) scenario.
There is a constant scale, namely the AdS curvature scale, to
which quantities can be compared.

The paper is organized as follows. In section 2, we explain the
general set up for the multibrane system. The junction conditions
for this system is derived for each brane by using geometrical
approach. In section 3, we give the basic formulation of the
gradient expansion method by solving five dimensional equations of
motion and imposing the Dirichlet boundary condition at the brane
position. We give the basic formulations of the extrinsic
curvature at the zeroth order and the first order expansions. The
first order solution of the effective equations of motion on the
brane is derived in section 4. In section 5, we address our result
into the scalar-tensor theory. We conclude our results in section
6.

\section{The general setup and background solution}

In this section we discuss three 3-brane embedded in a
five-dimensional space-time. This model is a straightforward
extension of the RS I model. In more detail, we have three
parallel 3-branes in an $AdS_5$ space with negative cosmological
constant. The fifth dimension has the geometry of an orbifold
$S^1/Z_2$ and the branes are located at $y=0$ (A-brane with
tension $\sigma_A$), $y=y_A$ (C-brane with tension $\sigma_C$) and
$y=y_B$ (B-brane with tension $\sigma_B$). Note that A-brane and
B-brane are placed at the fixed points of the orbifold $S^1/Z_2$
but C-brane is not. The region between two brane, namely region A
[$0,y_A$] and region B [$y_A,y_B$], is characterized by two
different values in two different slice of $AdS_5$ curvature
scales i.e., $l_A$ and $l_B$, respectively. Furthermore the
metrics on the three branes are all connected by an conformal
transformation, so in principle it is enough to derive the
four-dimensional effective equations of motion on one of the
branes. In the following we derive the four-dimensional Einstein
equations of motion on each branes. The action that describes this
configuration is
\begin{equation}\label{action}
    S = \frac{1}{2\kappa^2} \int d^5 x \sqrt{-g} \left(\mathcal{R} +
        \frac{12}{l^2_{A,B}}\right) + \sum_{i=A,B,C} \int d^4x
        \sqrt{-g^{i-brane}}(-\sigma_i + L^i_{matter}),
\end{equation}
where $\mathcal{R}$, $\kappa$ and $g$ are the scalar curvature,
the gravitational constant in 5-dimensions, respectively. The
metric $g^{i-brane}$ is the induced metric on the branes and
$\sigma_i$ their tensions. Notice that we have allowed the
curvature scales to have different values to either side of the
middle brane, represented by $l_A$ and $l_B$, which leads to
different metrics in the two bulk regions. The coordinate system
chosen is
\begin{equation}\label{ansatzmetric}
    ds^2=e^{2\phi(y,x)}dy^2 + g_{\mu\nu}(y,x)dx^\mu dx^\nu.
\end{equation}

The proper distance between two branes with fixed $x$ coordinates
can be written as
\begin{eqnarray}
    d_A(x) &=& \int_0^{y_A}dy'e^{\phi(y',x)} \ ,
    \label{properA} \\
    d_B(x) &=& \int_{y_A}^{y_B}dy'e^{\phi(y',x)} \ .
    \label{properB}
\end{eqnarray}
And the total proper distance is given by
\begin{equation}
    d_{tot}(x) = d_A(x) + d_B(x) = \int_0^{y_B}dy'e^{\phi(y',x)} \ .
    \label{totproper}
\end{equation}

The Einstein equations that arise from this coordinate system are:
\begin{eqnarray}
    & & e^{-\phi} (e^{-\phi} K_\mu^{\ \nu })_{,y}
            - (e^{-\phi} K)(e^{-\phi} K_\mu^{\ \nu})
            + {}^{(4)}R_\mu^\nu -D_\mu D^\nu \phi
            -D_\mu \phi  D^\nu \phi \nonumber \\
            && \qquad = - {4\over l^2_{A,B}} \delta_\mu^\nu
            + \kappa^2 \left({1\over 3} \delta_\mu^\nu \sigma_A
            +T^{A\nu}_{\mu} -{1\over 3} \delta_\mu^\nu T^A \right)
            e^{-\phi} \delta (y) \nonumber \\
            && \qquad  \qquad + {\kappa^2 \over 2} \left({1\over 3} \delta_\mu^\nu \sigma_C
            +\tilde{T}^{C\nu}_{\mu} -{1\over 3} \delta_\mu^\nu
            \tilde{T}^C
            \right) e^{-\phi} \delta (y-y_A) \nonumber \\
            && \qquad  \qquad + \kappa^2 \left({1\over 3} \delta_\mu^\nu
            \sigma_B
            +\hat{T}^{B\nu}_{\mu} -{1\over 3} \delta_\mu^\nu
            \hat{T}^B
            \right) e^{-\phi} \delta (y-y_B) \ ,
            \label{eq:k-munu}
\end{eqnarray}
and
\begin{eqnarray}
    & & e^{-\phi} (e^{-\phi} K )_{, y}
            - (e^{-\phi} K^{\alpha  \beta})(e^{-\phi} K_{\alpha  \beta} )
            - D^\alpha D_\alpha \phi
            -D^\alpha \phi D_\alpha \phi \nonumber \\
            &&\qquad     = - {4\over l^2_{A,B}}
            -{\kappa^2 \over 3} \left(-4\sigma_A +T^A \right)
            e^{-\phi}\delta (y)
            -{\kappa^2 \over 6} \left(-4\sigma_C +\tilde{T}^C \right)
            e^{-\phi}\delta (y-y_A) \nonumber \\
            && \qquad  \qquad  -{\kappa^2 \over 3} \left(-4\sigma_B +\hat{T}^B \right)
            e^{-\phi}\delta (y-y_B)  \ ,
            \label{eq:k-yy}
\end{eqnarray}
and
\begin{eqnarray}
    & & D_\nu (e^{-\phi} K_{\mu }^{\nu})
            - D_\mu (e^{-\phi} K )   =  0  \ ,
            \label{eq:k-ymu}
\end{eqnarray}
where the appropriate $AdS_5$ curvature scale is chosen for each
region. ${}^{(4)}R_{\mu}^{\nu}$ is the curvature on the brane and
$D_\mu$ denotes the covariant derivative with respect to the
metric $g_{\mu\nu}$. When the metric changes between regions of
the bulk, it is necessary to include a term in the boundary action
that depends on the trace of the extrinsic curvature.

The extrinsic curvature tensor is defined by $K_{\mu\nu} =
-g_{\mu\nu,y}/2$. For all three branes, the junction conditions
read
\begin{eqnarray}
    e^{-\phi} \left[ K_\mu^\nu - \delta_\mu^\nu K \right] |_{y=0}
            &=& {\kappa^2 \over 2}
            \left( -\sigma_A \delta_\mu^\nu
            + T^{A\nu}_{\mu} \right) \ ,
            \label{JunctionA}  \\
    e^{-\phi} \left[ K_\mu^\nu - \delta_\mu^\nu K \right]^{+}_{-} |_{y=y_A}
            &=& {\kappa^2}
            \left( -\sigma_C \delta_\mu^\nu
            + \tilde{T}^{C\nu}_{\mu} \right) \ ,
            \label{JunctionC}   \\
    e^{-\phi} \left[ K_\mu^\nu - \delta_\mu^\nu  K \right] |_{y=y_B}
            &=& -{\kappa^2 \over 2}
            \left( -\sigma_B \delta_\mu^\nu
            + \hat{T}^{B\nu}_{\mu} \right) \ ,
            \label{JunctionB}
\end{eqnarray}
where we have used that A-brane and B-brane follow a $Z_2$
symmetry but C-brane is no $Z_2$ symmetry. Here, we assume that
the direction of the normal vector field to a brane is chosen to
be the same all way through the bulk for all the three brane.
Decompose the extrinsic curvature into the traceless part and the
trace part
\begin{equation}
    e^{-\phi}K_{\mu\nu}
            = \Sigma_{\mu\nu} + {1\over 4} g_{\mu\nu} Q  \ , \quad
        Q = - e^{-\phi}{\partial \over \partial y}\log \sqrt{-g}    \ ,
        \label{eq:decompose}
\end{equation}
then, we obtain the basic equations;
\begin{eqnarray}
    & & e^{-\phi} \Sigma^\nu_{\mu , y} - Q \Sigma_\mu^{\nu}
            = -\Bigg[ {}^{(4)}R_\mu^{\nu}  - {1\over 4} \delta_\mu^\nu {}^{(4)}R
                -D_\mu D^\nu \phi
                -D_\mu \phi  D^\nu \phi \nonumber \\
                && \qquad \qquad \qquad \qquad \qquad + {1\over 4} \delta_\mu^\nu
                \left( D^\alpha D_\alpha \phi
            +D^\alpha \phi D_\alpha \phi \right)
            \Bigg]      \ ,
            \label{eq:munu-traceless} \\
    & & {3\over 4} Q^2 - \Sigma^\alpha_{\ \beta} \Sigma^\beta_{\ \alpha}
            = \left[ {}^{(4)}R \right] + {12\over l^2_{A,B}}   \ ,
            \label{eq:munu-trace} \\
    & & e^{-\phi}  Q_{, y} -{1\over 4}Q^2
            - \Sigma^{\alpha \beta} \Sigma_{\alpha \beta}
            =D^\alpha D_\alpha \phi
            + D^\alpha \phi D_\alpha \phi - {4\over l^2_{A,B}}  \ ,
            \label{eq:yy}  \\
    & & D_\nu \Sigma_{\mu }^{\nu}
            - {3\over 4} D_\mu Q = 0   \ .
            \label{eq:ymu}
\end{eqnarray}
And the junction conditions read
\begin{eqnarray}
    \left[ \Sigma_{\mu}^{\nu}
        - {3\over 4} \delta_\mu^\nu Q \right] \Bigg|_{y=0}
            &=& {\kappa^2 \over 2}  \left( -\sigma_A \delta_\mu^\nu
            + T^{A\nu}_{\mu} \right)  \ ,
            \label{JunctionA1} \\
    \left[ \Sigma_{\mu}^{\nu}
        - {3\over 4} \delta_{\mu}^{\nu} Q \right]^{+}_{-} \Bigg|_{y=y_A}
            &=& {\kappa^2} \left( -\sigma_C \delta_{\mu}^{\nu} +
            \tilde{T}^{C\nu}_{\mu} \right)    \ ,
            \label{JunctionC1} \\
    \left[ \Sigma_{\mu}^{\nu}
        - {3\over 4} \delta_{\mu}^{\nu} Q \right] \Bigg|_{y=y_B}
            &=& -{\kappa^2 \over 2} \left( -\sigma_B \delta_{\mu}^{\nu} +
            \hat{T}^{B\nu}_{\mu} \right)    \ .
            \label{JunctionB1}
\end{eqnarray}
The notation $[X]^{+}_{-}$ indicates that we evaluate the quantity
$X$ on both sides of the brane and take difference, $[X]^{+}_{-} =
X^{+} - X_{-}$.

\section{The gradient expansion method}
In this section we derive the effective equations of motion using
the low energy expansion method (gradient expansion method) first
proposed by Kanno and Soda~\cite{KS1} to study our scenario. In
this method, the full five dimensional equations of motion are
solved, at different orders, in the bulk by performing a
perturbation expansion in the metric. The parameter of expansion
is defined so that the low energy regime corresponds to a regime
in which the energy density ($\rho $) on the brane is smaller then
the brane tension ($\sigma$), $\rho << \sigma$. In this regime,
the parameter of expansion can be expressed as
\begin{equation}
    \epsilon = \left( {l \over L} \right)^2 \ ,
    \label{Par:expansion}
\end{equation}
where $l$ is the bulk curvature scale of the $AdS_5$ and $L$ is
the brane curvature scale. According to the parameter of expansion
(\ref{Par:expansion}), the quantities $\Sigma_\mu^{\nu}$ are
expanded as
\begin{equation}
    \Sigma_\mu^{\ \nu} = \Sigma^{(0)\nu}_{\mu}
    +\Sigma^{(1)\nu}_{\mu} + \Sigma^{(2)\nu}_{\mu}
    +\cdots  \ .
    \label{expansion:quantity}
\end{equation}
Then, the iteration scheme consists in writing the metric
$g_{\mu\nu}$ as a sum of local tensors built out of the induced
metric on the brane,
\begin{eqnarray}
        &&  g_{\mu\nu} (y,x^\mu ) =
        a^2 (y,x) \left[ h_{\mu\nu} (x^\mu)
        + g^{(1)}_{\mu\nu} (y,x^\mu)
            + g^{(2)}_{\mu\nu} (y, x^\mu ) + \cdots  \right]  \ ,
            \label{expansion:metric} \\
    &&  g^{(n)}_{\mu\nu} (y=\bar{y} ,x^\mu ) =  0    \ , \quad n=1,2,3,...
        \label{BC}
\end{eqnarray}
Here, $\bar{y}$ is a generic point in which the Dirichlet
condition is taken on the brane.

In the following, we derive the zeroth order and first order
solutions using the above scheme.

\subsection{Zeroth order solutions}

At zeroth order matter is neglected, we intend vacuum brane, and
going at higher orders means we are considering perturbation of
the vacuum solution as matter is added to the brane. The equations
to solve are
\begin{eqnarray}
    & & e^{-\phi} \Sigma^{(0)\nu}_{\mu , y} - Q^{(0)} \Sigma_\mu^{(0)\nu}
            = 0
            \label{munu-traceless:zeroorder} \\
    & & {3\over 4} Q^{(0)2} - \Sigma^{(0)\alpha}_{\beta} \Sigma^{(0)\beta}_{\alpha}
            = {12\over l^2_{A,B}}   \ ,
            \label{munu-trace:zeroorder} \\
    & & e^{-\phi}  Q^{(0)}_{, y} -{1\over 4}Q^{(0)2}
            - \Sigma^{(0)\alpha \beta} \Sigma^{(0)}_{\alpha \beta}
            = - {4\over l^2_{A,B}}  \ ,
            \label{yy:zeroorder}  \\
    & & D_\nu \Sigma_{\mu }^{(0)\nu}
            - {3\over 4} D_\mu Q^{(0)} = 0   \ .
            \label{ymu:zeroorder}
\end{eqnarray}
The junction conditions are given by
\begin{eqnarray}
    \left[ \Sigma_{\mu}^{(0)\nu}
        - {3\over 4} \delta_\mu^\nu Q^{(0)} \right] \Bigg|_{y=0}
            &=& - {\kappa^2 \over 2} \sigma_A \delta_\mu^\nu \ ,
            \label{JunctionA1:zero} \\
    \left[ \Sigma_{\mu}^{(0)\nu}
        - {3\over 4} \delta_{\mu}^{\nu} Q^{(0)} \right]^{+}_{-} \Bigg|_{y=y_A}
            &=& -{\kappa^2} \sigma_C \delta_{\mu}^{\nu}   \ .
            \label{JunctionC1:zero} \\
    \left[ \Sigma_{\mu}^{(0)\nu}
        - {3\over 4} \delta_{\mu}^{\nu} Q^{(0)} \right] \Bigg|_{y=y_B}
            &=& {\kappa^2 \over 2} \sigma_B \delta_{\mu}^{\nu}  \ .
            \label{JunctionB1:zero}
\end{eqnarray}
Integrating equation (\ref{munu-traceless:zeroorder}) and using
the constraint (\ref{ymu:zeroorder}) we obtain the solution of the
traceless part of the extrinsic curvature at zeroth order
\begin{equation}
        \Sigma_{\mu}^{(0)\nu} \Big|_{-} =\Sigma_{\mu}^{(0)\nu} \Big|_{+}= 0 \ .
        \label{sol:munutraceless}
\end{equation}
And the trace part of the extrinsic curvature
\begin{equation}
        Q^{(0)} \Big|_{-}= {4 \over l_{A}}, \qquad  Q^{(0)} \Big|_{+} = {4 \over l_{B}} \ ,
        \label{sol:munutrace}
\end{equation}
where we have inserted (\ref{sol:munutraceless}) into
(\ref{munu-trace:zeroorder}) and "$|_{\pm}$" denotes the solutions
of each region of the bulk space-time. Inserting
(\ref{sol:munutraceless}) and (\ref{sol:munutrace}) into
(\ref{eq:decompose}), the extrinsic curvature at zeroth order is
\begin{equation}
        K^{(0)}_{\mu\nu}\Big|_{\pm} =  {e^{\phi} \over l_{A,B}} g^{(0)}_{\mu\nu}\Big|_{\pm} \ .
        \label{sol:ECdecompse}
\end{equation}
In this order, the equation (\ref{sol:ECdecompse}) gives the
evolution of the extrinsic curvature in different regions of the
bulk space-time which correspond to two different values of the
$AdS$ curvature scales, $l_A$ and $l_B$.

From definition of the extrinsic curvature and using equation
(\ref{sol:ECdecompse}) we need to solve two different solutions of
the metric
\begin{eqnarray}
    - {1 \over 2} {\partial \over \partial
    y}g^{(0)}_{\mu\nu}\Big|_{-} &=&  {e^{\phi} \over l_{A}}
    g^{(0)}_{\mu\nu}\Big|_{-} \ ,  \qquad  0 < y < y_A \ ,
    \label{solmetric:zero order minus} \\
    - {1 \over 2} {\partial \over \partial
    y}g^{(0)}_{\mu\nu}\Big|_{+} &=&  {e^{\phi} \over l_{A}}
    g^{(0)}_{\mu\nu}\Big|_{+} \ , \qquad y_A < y < y_B \ .
    \label{solmetric:zero order plus}
\end{eqnarray}
Integrating equations (\ref{solmetric:zero order minus}) and
(\ref{solmetric:zero order plus}) we get the zeroth order metric
as
\begin{equation}
    ds^2=e^{2\phi(y,x)}dy^2 + a^2_{\mp}(y,x) h^{\mp}_{\mu\nu}(x)dx^\mu dx^\nu \ ,
    \label{Sol:order-0 ansatzmetric}
\end{equation}
where
\begin{eqnarray}
    a_{-} (y,x) &=& e^{ - d_A (y,x)/ l_A}   \ ,
    \label{solmet:A} \\
    a_{+} (y,x) &=& e^{d_B (y,x) / l_B}   \ .
    \label{solmet:B}
\end{eqnarray}
Here, we have integrated from a generic point $y'$ to $y$ such
that $d_{A,B} = \int_{y'}^y e^{\phi(y,x)}dy$ is the proper
distance between a generic point and $y$. The tensor
$h^{\mp}_{\mu\nu}$ is the induced metric tensor depending on the
brane coordinates. In fact, the factor $a_{\pm}$ is a conformal
factor that relates the metric on the branes. In this case we have
\begin{eqnarray}
    h_{\mu\nu}^{C-} &=&  a_{-}^2 h_{\mu\nu}^{A}  \ ,
    \label{conf:A} \\
    h_{\mu\nu}^{C+} &=&  a_{+}^2  h_{\mu\nu}^{B}   \ .
    \label{conf:B}
\end{eqnarray}
where $h_{\mu\nu}^{C\mp}=g_{\mu\nu}^{C-brane}$,
$h_{\mu\nu}^{A}=g_{\mu\nu}^{A-brane}$ and
$h_{\mu\nu}^{B}=g_{\mu\nu}^{B-brane}$ are the induced metrics on
the C-brane, A-brane and B-brane, respectively.

From the junction conditions (\ref{JunctionA1:zero}) and
(\ref{JunctionB1:zero}) we have the fine tuning conditions for
A-brane tension and B-brane tension respectively,
\begin{eqnarray}
    \kappa^2 \sigma_A = {6 \over l_A} \ ,
    \label{0:JCA}\\
    \kappa^2 \sigma_B = -{6 \over l_B} \ .
    \label{0:JCB}
\end{eqnarray}
Furthermore, the junction condition (\ref{JunctionC1:zero}) yields
\begin{equation}
    \kappa^2 \sigma_C = - {3\over l_A} \left( {{\alpha -1} \over \alpha}  \right) \ .
    \label{0:JCC}
\end{equation}
where we have defined $\alpha = {l_B}/{l_A}$. This equation is a
fine tuning condition for C-brane tension in terms of the
curvature scales $l_A$ and $l_B$ of the slices of $AdS_5$ bulk.
The equation (\ref{0:JCA}) - (\ref{0:JCC}) also implies the
relation for the brane tensions:
\begin{equation}
    \sigma_A + 2 \sigma_C + \sigma_B = 0 \ .
    \label{0:tension}
\end{equation}
The RS I model is obtained for $\alpha = 1$ where C-brane is
absent, $\sigma_C = 0$. For $\alpha < 1$ we have $\sigma_C > 0$
which correspond to inflation C-brane~\cite{KSW}. Various brane
world models can be recovered by using relations (\ref{0:JCC}) and
(\ref{0:tension})~\cite{IMPRS, GRS, LR, PRZ}.

\subsection{First order solutions}

The aim of this subsection is now to solve the four basic
equations (\ref{eq:munu-traceless}) - (\ref{eq:ymu}) at first
order. In this order, the solution can be obtained by taking into
account the terms neglected at the zeroth order. We have
\begin{eqnarray}
    & & e^{-\phi} \Sigma^{(1)\nu}_{\mu , y} - {4 \over l_{A,B}} \Sigma_\mu^{(1)\nu}
            = -\Bigg[ {}^{(4)}R_\mu^{\nu}  - {1\over 4} \delta_\mu^\nu {}^{(4)}R
                -D_\mu D^\nu \phi - D_\mu \phi  D^\nu \phi \nonumber \\
                && \qquad \qquad \qquad \qquad \qquad + {1\over 4} \delta_\mu^\nu
                \left( D^\alpha D_\alpha \phi
            +D^\alpha \phi D_\alpha \phi \right)
            \Bigg]^{(1)}      \ ,
            \label{eq:munu-traceless1} \\
    & & {6 \over l_{A,B} } Q^{(1)}  = \left[ {}^{(4)}R \right]   \ ,
            \label{eq:munu-trace1} \\
    & & e^{-\phi}  Q^{(1)}_{, y} -{2\over l_{A,B}}Q^{(1)}
            = \left[ D^\alpha D_\alpha \phi
            + D^\alpha \phi D_\alpha \phi \right]^{(1)}   \ ,
            \label{eq:yy1}  \\
    & & D_\nu \Sigma_{\mu }^{(1) \nu}
            - {3\over 4} D_\mu Q^{(1)} = 0   \ ,
            \label{eq:ymu1}
\end{eqnarray}
where the superscript $(1)$ represents the order of the gradient
expansion. The junction conditions are given by
\begin{eqnarray}
    \left[ \Sigma_{\mu}^{(1)\nu}
        - {3\over 4} \delta_\mu^\nu Q^{(1)} \right] \Bigg|_{y=0}
            &=& {\kappa^2 \over 2} T^{A\nu}_{\mu}   \ ,
            \label{JunctionA:1-O} \\
    \left[ \Sigma_{\mu}^{(1)\nu}
        - {3\over 4} \delta_{\mu}^{\nu} Q^{(1)} \right]^{+}_{-} \Bigg|_{y=y_A}
            &=& {\kappa^2}\tilde{T}^{C\nu}_{\mu}    \ .
            \label{JunctionC1:1-O} \\
    \left[ \Sigma_{\mu}^{(1)\nu}
        - {3\over 4} \delta_{\mu}^{\nu} Q^{(1)} \right] \Bigg|_{y=y_B}
            &=& -{\kappa^2 \over 2}\hat{T}^{B\nu}_{\mu}    \ .
            \label{JunctionB1:1-O}
\end{eqnarray}
Here, $T^{A\nu}_{\mu}$, $\tilde{T}^{C\nu}_{\mu} $ and
$\hat{T}^{B\nu}_{\mu}$ are the energy-momentum tensors with the
indices raised by the induced metric on the A-brane, C-brane and
B-brane, respectively. We compute the Ricci tensor $\left[
{}^{(4)}R_{\mu}^{\nu}\right]^{(1)}$ of a metric
$a^2_{\mp}h_{\mu\nu}^{\mp}$. The Christoffel symbol is
\begin{equation}
    \Gamma^\rho_{\mu\nu}(g) = \Gamma^\rho_{\mu\nu}(h) +
    D_\mu \ln a \delta^\rho_\nu + D_\nu \ln a
    \delta^\rho_\mu - h_{\mu\nu} D^\rho \ln a \ .
    \label{Chris: Symbol}
\end{equation}
Here $D_\mu$ is a covariant derivative with respect to the metric
$h_{\mu\nu}$. Using equation (\ref{Chris: Symbol}) the Ricci
tensor of a metric $g_{\mu\nu}=a^2_{-}h_{\mu\nu}^{-}$ is given by
\begin{eqnarray}
    \left[ {}^{(4)}R_{\mu}^{\nu}(g)\right]^{(1)}_{-} &=& {1\over a^2_{-}} \Bigg[ {}^{(4)}R_{\mu}^{\nu}(h^{-})
    + {2\over l_{A}} \Bigg( \mathcal{D}_\mu \mathcal{D}^\nu d_{A} + {1\over l_{A}} \mathcal{D}_\mu d_{A} \mathcal{D}^\nu d_{A}
     \Bigg) \nonumber\\
     && + {1\over l_{A}} \delta_{\mu}^{\nu} \Bigg(\mathcal{D}_\sigma \mathcal{D}^\sigma d_{A}
     -  {2\over l_{A}}\mathcal{D}_\sigma d_{A} \mathcal{D}^\sigma d_{A}\Bigg) \Bigg] \ .
    \label{Ricci:neg}
\end{eqnarray}
Contracting indices $\mu$ and $\nu$ of the equation
(\ref{Ricci:neg}), we obtain the expression for the Ricci scalar
\begin{equation}
    \left[ {}^{(4)}R(g)\right]^{(1)}_{-} = {1\over
    a^2_{-}} \Bigg[ {}^{(4)}R(h^{-})
    + {6\over l_{A}} \Bigg( \mathcal{D}_\sigma \mathcal{D}^\sigma d_{A}
     -  {1\over l_{A}}\mathcal{D}_\sigma d_{A} \mathcal{D}^\sigma d_{A}\Bigg) \Bigg] \ .
    \label{Ricciscalar:neg}
\end{equation}
On the other hand, for the Ricci tensor and Ricci scalar of a
metric $g_{\mu\nu}=a^2_{+}h_{\mu\nu}^{+}$, respectively are given
by
\begin{eqnarray}
    \left[ {}^{(4)}R_{\mu}^{\nu}(g)\right]^{(1)}_{+} &=& {1\over a^2_{+}} \Bigg[ {}^{(4)}R_{\mu}^{\nu}(h^{+})
    - {2\over l_{B}} \Bigg( \mathcal{D}_\mu \mathcal{D}^\nu d_{B} - {1\over l_{B}} \mathcal{D}_\mu d_{B} \mathcal{D}^\nu d_{B}
     \Bigg) \nonumber\\
     && - {1\over l_{B}} \delta_{\mu}^{\nu} \Bigg(\mathcal{D}_\sigma \mathcal{D}^\sigma d_{B}
     +  {2\over l_{B}}\mathcal{D}_\sigma d_{B} \mathcal{D}^\sigma d_{B}\Bigg) \Bigg] \ ,
    \label{Ricci:pos} \\
    \left[ {}^{(4)}R(g)\right]^{(1)}_{+} &=& {1\over
    a^2_{+}} \Bigg[ {}^{(4)}R(h^{+})
    - {6\over l_{B}} \Bigg( \mathcal{D}_\sigma \mathcal{D}^\sigma d_{B}
    +  {1\over l_{B}}\mathcal{D}_\sigma d_{B} \mathcal{D}^\sigma d_{B}\Bigg) \Bigg] \ .
    \label{Ricciscalar:pos}
\end{eqnarray}

We also express the kinetic terms of $\phi$ in terms of the proper
distance, the indices $-$ and $+$ have been omitted to give a
general case, as follows
\begin{eqnarray}
     & & \Bigg[ -\mathcal{D}_\mu \mathcal{D}^\nu \phi - \mathcal{D}_\mu \phi  \mathcal{D}^\nu \phi + {1\over 4} \delta_\mu^\nu
            \Bigg( \mathcal{D}^\alpha \mathcal{D}_\alpha \phi
            +\mathcal{D}^\alpha \phi \mathcal{D}_\alpha \phi \Bigg)
            \Bigg]^{(1)} \nonumber \\
            & & \qquad =  {e^{-\phi}\over a^2} {\partial \over \partial y}
            \Bigg[\Bigg( \mathcal{D}_\mu \mathcal{D}^\nu d
            - {1\over 4} \delta_\mu^\nu \mathcal{D}_\sigma \mathcal{D}^\sigma d \Bigg) - {1\over l}\Bigg( \mathcal{D}_\mu d \mathcal{D}^\nu d
            - {1\over 4} \delta_\mu^\nu \mathcal{D}_\sigma d \mathcal{D}^\sigma d \Bigg) \Bigg] \ .
            \label{eq:munuphi}
\end{eqnarray}

In the equations above, $\mathcal{D}_\mu$ denotes the covariant
derivative with respect to the induced metric $h_{\mu\nu}^{\mp}$
on the brane. And we have written the Ricci tensor in terms of the
proper distance for a generic point on the y-axis.

Substituting (\ref{Ricciscalar:neg}) into (\ref{eq:munu-trace1})
we obtain the solutions of negative side of the C-brane as
\begin{equation}
    Q^{(1)}|_{-} = {l_{A}\over a^2_{-}} \Bigg[ {1\over 6} {}^{(4)}R +
    {1\over l_{A}} \Bigg( \mathcal{D}_\sigma \mathcal{D}^\sigma d_{A}
     - {1\over l_{A}} \mathcal{D}_\sigma d_{A} \mathcal{D}^\sigma d_{A}
     \Bigg)\Bigg] \ .
     \label{soltrace:neg}
\end{equation}
We are now to obtain the traceless part of the extrinsic curvature
at first order $\Sigma_{\mu}^{(1)\nu}$. Substituting
(\ref{Ricciscalar:neg}) and (\ref{eq:munuphi}) into
(\ref{eq:munu-traceless1}) and integrating we obtain
\begin{eqnarray}
    \Sigma_{\mu}^{(1)\nu}(y,x) |_{-} &=&
    {l_A \over 2 a^2_{-}} \left( {}^{(4)}R_\mu^\nu - {1\over 4}
    \delta_\mu^\nu {}^{(4)}R \right) + {1\over a^2_A} \left( \mathcal{D}_\mu \mathcal{D}^\nu
    d_A - {1\over 4} \delta_\mu^\nu \mathcal{D}_\sigma \mathcal{D}^\sigma
    d_A \right) \nonumber \\
    && {1\over l_A a^2_{-} } \left( \mathcal{D}_\mu d_A \mathcal{D}^\nu
    d_A - {1\over 4} \delta_\mu^\nu \mathcal{D}_\sigma d_A
    \mathcal{D}^\sigma d_A \right)
    + {\chi_\mu^{\nu}|_{-} \over a^4_A} \ ,
    \label{traceless:neg}
\end{eqnarray}
where $\chi_\mu^{\nu}|_{-}$ is a integration constant which
satisfy $\chi_\mu^{\mu}|_{-}=0$ and  $\mathcal{D}_\nu
\chi_\mu^{\nu}|_{-}=0$.

And the solutions of positive side of the C-brane are
\begin{eqnarray}
    Q^{(1)}|_{+} &=& {l_{B}\over a^2_{+}} \Bigg[ {1\over 6} {}^{(4)}R
    - {1\over l_{B}} \Bigg( \mathcal{D}_\sigma \mathcal{D}^\sigma d_{B}
     - {1\over l_{B}} \mathcal{D}_\sigma d_{B} \mathcal{D}^\sigma d_{B}
     \Bigg)\Bigg] \ ,
     \label{soltrace:pos}\\
    \Sigma_{\mu}^{(1)\nu}(y,x) |_{+} &=&
    {l_B \over 2 a^2_{+}} \left( {}^{(4)}R_\mu^\nu - {1\over 4}
    \delta_\mu^\nu {}^{(4)}R \right) - {1\over a^2_B} \left( \mathcal{D}_\mu \mathcal{D}^\nu
    d_B - {1\over 4} \delta_\mu^\nu \mathcal{D}_\sigma \mathcal{D}^\sigma
    d_B \right) \nonumber \\
    && {1\over l_B a^2_{+} } \left( \mathcal{D}_\mu d_B \mathcal{D}^\nu
    d_B - {1\over 4} \delta_\mu^\nu \mathcal{D}_\sigma d_B
    \mathcal{D}^\sigma d_B \right)
    + {\chi_\mu^{\nu}|_{+} \over a^4_B} \ .
    \label{traceless:pos}
\end{eqnarray}
where $\chi_\mu^{\nu}|_{+}$ is a integration constant which
satisfy $\chi_\mu^{\mu}|_{+}=0$ and  $\mathcal{D}_\nu
\chi_\mu^{\nu}|_{+}=0$.

The integration constants $\chi_\mu^{\nu}|_{-}$ and
$\chi_\mu^{\nu}|_{+}$ are non-local terms , which corresponds to
the projection on the brane of the five- dimensional Weyl tensor.
Therefore they carry the information of the bulk gravitational
fields. In the following subsection we give explicitly the
expression of these terms.

\section{The effective equations of motion on the branes}

In this section,  we derive the effective equations of motion on
each brane, at the first order, by substituting the equations of
the extrinsic curvature into the junction conditions
(\ref{JunctionC1:1-O})-(\ref{JunctionB1:1-O}). After imposing the
trace of the projective Weyl tensor vanishes, $\chi_\mu^\mu =0$,
we then obtain the equations of motion for the scalar (radion)
fields. We start to derive the equations of motion on the middle
brane.

\subsection{Solutions on the middle brane}

At the middle brane (C-brane) is no $Z_2$ symmetry, the traceless
part of the extrinsic curvature is asymmetric. Therefore it has
two values in different sides of C-brane. Using equation
(\ref{soltrace:neg}) - (\ref{traceless:pos}), the junction
condition at C-brane is written as
\begin{equation}
    {\chi_\mu^{\nu}|_{+} \over e^{4d_B /l_A}} - {\chi_\mu^{\nu}|_{-} \over
    e^{-4d_A/l_A}} +  {l_A \over 2} \left( \alpha - 1 \right)  G_\mu^\nu (h^C) = \kappa^2 T_\mu^{C\nu} \ .
    \label{JC: atC}
\end{equation}
where we have used the Einstein equations on the negative and
positive sides at C-brane, $G_\mu^\nu|_{-}=G_\mu^\nu|_{+} \equiv
G_\mu^\nu (h^C)$. The junction condition at A-brane is given by
\begin{eqnarray}
    & & \chi_\mu^{\nu}|_{-} + {l_A \over 2 e^{2d_A/l_A}}G_\mu^\nu (h^C)
        - {1 \over  e^{2d_A/l_A}} \left( \mathcal{D}_\mu \mathcal{D}^\nu d_A
        - \delta_\mu^\nu \mathcal{D}_\sigma \mathcal{D}^\sigma d_A \right) \nonumber \\
        && \qquad \qquad \qquad + {1 \over  l_A e^{2d_A/l_A} } \left( \mathcal{D}_\mu d_A \mathcal{D}^\nu d_A
        + {1\over 2} \delta_\mu^\nu \mathcal{D}_\sigma d_A \mathcal{D}^\sigma d_A
        \right) = {\kappa^2 \over 2 e^{2d_A/l_A}} T_\mu^{A\nu} \ ,
        \label{JC: atA}
\end{eqnarray}
and the junction condition at B-brane yields
\begin{eqnarray}
    & & \chi_\mu^{\nu}|_{+} + {\alpha l_A \over 2 }e^{2d_B/l_B}G_\mu^\nu(h^C)
        + e^{2d_B/l_B} \left( \mathcal{D}_\mu \mathcal{D}^\nu d_B
        - \delta_\mu^\nu \mathcal{D}_\sigma \mathcal{D}^\sigma d_B \right) \nonumber \\
        && \qquad \qquad + {1 \over  l_B} e^{2d_B/l_B} \left( \mathcal{D}_\mu d_B \mathcal{D}^\nu d_B
        + {1\over 2} \delta_\mu^\nu \mathcal{D}_\sigma d_B \mathcal{D}^\sigma d_B
        \right) = - {\kappa^2 \over 2 }e^{2d_B/l_B} T_\mu^{B\nu} \ ,
        \label{JC: atB}
\end{eqnarray}
where $\mathcal{D}_\mu$ is the derivative covariant with respect
to the induced metric on the C-brane. To get equations (\ref{JC:
atA}) and (\ref{JC: atB}), we have used the conformal
transformation of the metric both at negative side of C-brane and
A-brane and at positive side of C-brane and B-brane as follows
\begin{eqnarray}
    h^{C-}_{\mu\nu} &=& e^{-2d_A/l_A} h^A_{\mu\nu} \ , \\
    h^{C+}_{\mu\nu} &=& e^{2d_B/l_B} h^B_{\mu\nu} \ .
\end{eqnarray}
Therefore, the index of $T_\mu^{A\nu}$ and $T_\mu^{B\nu}$ are the
energy-momentum tensors with the indices raised by the induced
metric on the both side of C-brane, while $\tilde{T}_\mu^{A\nu}$
and $\hat{T}_\mu^{B\nu}$ are he energy-momentum tensors with the
indices raised by the induced metric on the A-brane and B-brane,
respectively. In order to obtain the effective equations of motion
at C-brane, one can subtract equation (\ref{JC: atA}) with respect
to (\ref{JC: atB}). Then, substituting this result into equation
(\ref{JC: atC}) yields
\begin{eqnarray}
    G_\mu^\nu (h^C)&=&
    {{2 \kappa^2}\over l_A( \Phi_C + \alpha \Psi_C) } \left[ T_\mu^{C\nu} + {1 \over 2}(1 + \Phi_C) T_\mu^{A\nu} + {1 \over 2}(1 -
    \Psi_C)
    T_\mu^{B\nu}  \right] \nonumber \\
    && + {1 \over ( \Phi_C + \alpha \Psi_C) } \Bigg[ \left( \mathcal{D}_\mu \mathcal{D}^\nu \Phi_C  - \delta_\mu^{\nu} \mathcal{D}_\sigma \mathcal{D}^\sigma \Phi_C \right)\nonumber\\
    && + {\omega (\Phi_C) \over \Phi_C} \left( \mathcal{D}_\mu \Phi_C \mathcal{D}^\nu \Phi_C
    - {1\over 2} \delta_\mu^{\nu} \mathcal{D}_\sigma \Phi_C \mathcal{D}^\sigma \Phi_C \right) \Bigg] \nonumber \\
    &&  + {\alpha  \over ( \Phi_C + \alpha \Psi_C)} \Bigg[ \left( \mathcal{D}_\mu \mathcal{D}^\nu \Psi_C -
    \delta_\mu^{\nu} \mathcal{D}_\sigma \mathcal{D}^\sigma \Psi_C \right) \nonumber \\
    &&  + {\omega (\Psi_C) \over \Psi_C} \left( \mathcal{D}_\mu \Psi_C \mathcal{D}^\nu \Psi_C
    - {1\over 2} \delta_\mu^{\nu}  \mathcal{D}_\sigma \Psi_C \mathcal{D}^\sigma \Psi_C  \right) \Bigg],
    \label{effEq:C}
\end{eqnarray}
where we have defined two scalar fields
\begin{eqnarray}
    \Phi_C &=& e^{2d_A/l_A} -1, \qquad~~~\qquad \Psi_C = 1 - e^{-2d_B/l_B} \ ,
    \label{def:Phi} \\
    \omega (\Phi_C) &=& - {3\over 2} {\Phi_C \over (1 + \Phi_C )}, \qquad
     \omega (\Psi_C) = {3\over 2} {\Psi_C \over (1 + \Psi_C )} \ .
     \label{def:omega}
\end{eqnarray}
Inserting (\ref{effEq:C}) into (\ref{JC: atA}) and (\ref{JC:
atB}), respectively, we obtain
\begin{eqnarray}
    \chi_\mu^{\nu}|_{-} &=& -{{\kappa^2 (1+\Phi_C)} \over (\Phi_C + \alpha \Psi_C)}
    \left[ T_\mu^{C\nu} + {1\over 2 }(1 - \alpha\Psi_C) T_\mu^{A\nu}+{1\over 2}(1-\Psi_C) T_\mu^{B\nu} \right] \nonumber \\
    &&-{l_A \over 2(\Phi_C + \alpha \Psi_C)} \left[ \left( 1 - \alpha \Psi_C \right) P_\mu^{\nu} (\Phi_C)
    + {\alpha^2} \left( 1 + \Phi_C \right) P_\mu^{\nu} (\Psi_C) \right]\ ,
    \label{chimin} \\
    \chi_\mu^{\nu}|_{+} &=& -{{\alpha \kappa^2 (1-\Psi_C)} \over ( \Phi_C + \alpha \Psi_C)}
    \left[ T_\mu^{C\nu} +  {1\over 2 }(1 +\Phi_C) T_\mu^{A\nu} + {1\over2}\left({\alpha  + \Phi_C \over \alpha }\right) T_\mu^{B\nu} \right] \nonumber \\
    && - {\alpha l_A \over 2(\Phi_C + \alpha \Psi_C)} \left[ \left( 1-\Psi_C \right) P_\mu^{\nu} (\Phi_C)
    +  \left( \alpha  +  \Phi_C \right) P_\mu^{\nu} (\Psi_C) \right] \ ,
    \label{chiplus}
\end{eqnarray}
where
\begin{eqnarray}
    P_\mu^{\nu} (\Phi_C) &=&\left( \mathcal{D}_\mu \mathcal{D}^\nu \Phi_C
    - \delta_\mu^{\nu} \mathcal{D}_\sigma \mathcal{D}^\sigma\Phi_C \right) \nonumber\\
    &&+ {\omega (\Phi_C) \over \Phi_C} \left( \mathcal{D}_\mu \Phi_C \mathcal{D}^\nu \Phi_C
    - {1\over 2} \delta_\mu^{\nu} \mathcal{D}_\sigma \Phi_C \mathcal{D}^\sigma \Phi_C \right) \ ,
    \label{Pmunu:Phi} \\
    P_\mu^{\nu} (\Psi_C) &=&\left( \mathcal{D}_\mu \mathcal{D}^\nu
    \Psi_C  - \delta_\mu^{\nu} \mathcal{D}_\sigma  \mathcal{D}^\sigma \Psi_C \right)\nonumber\\
    && + {\omega (\Psi_C) \over \Psi_C} \left( \mathcal{D}_\mu \Psi_C \mathcal{D}^\nu \Psi_C
     - {1\over 2} \delta_\mu^{\nu} \mathcal{D}_\sigma \Psi_C \mathcal{D}^\sigma \Psi_C \right) \ .
    \label{Pmunu:Psi}
\end{eqnarray}
The equations (\ref{chimin}) and (\ref{chiplus}) correspond with
discontinuity of the evolution for Weyl tensor in each regions.
The value $\chi_\mu^{\nu}|_{-}$ on the negative side corresponds
to the evolution toward A-brane and the value
$\chi_\mu^{\nu}|_{+}$ on the positive side corresponds to the
evolution toward B-brane.

The effective equations of motion for both scalar fields $\Phi_C$
and $\Psi_C$ can be obtained by using the condition
$\chi_\mu^{\mu}|_{-} =0$ and $\chi_\mu^{\mu}|_{+} =0$,
respectively. Then we get
\begin{eqnarray}
    \square \Phi_C &=& { \kappa^2 \over {l_A( 1- \alpha )}}\left[ {{(1 -
    \alpha ) T^A + 2  T^C} \over {2\omega(\Phi_C) +3 }} \right]   \nonumber\\
    &&- {1 \over {2\omega(\Phi_C) +3 }} {{d\omega(\Phi_C)} \over d\Phi_C}\mathcal{D}_\sigma \Phi_C \mathcal{D}^\sigma \Phi_C \ ,
    \label{eom:Phi} \\
    \square \Psi_C &=& { \kappa^2 \over {\alpha l_A(\alpha  - 1)}}\left[ {{(\alpha  -
    1) T^B + 2\alpha T^C} \over {2\omega(\Psi_C) +3 }} \right]\nonumber\\
    && - {1 \over {2\omega(\Psi_C) +3 }} {{d\omega(\Psi_C)} \over d\Psi_C}\mathcal{D}_\sigma \Psi_C \mathcal{D}^\sigma \Psi_C \ .
    \label{eom:Psi}
\end{eqnarray}
We see that the effective equations of motion for the scalar
fields dependent on the matter sources for two branes.

\subsection{Solutions on the other branes}

We now proceed to find the equations of motion on the other
branes. First, we derive the equations of motion on the A-brane.
Using the fact that the metric induced on the A-brane is related
by $h^A_{\mu\nu} = e^{2d_A\l_A}h^{C-}_{\mu\nu}$, the junction
condition on the A-brane is given by
\begin{equation}
    \chi_\mu^{\nu}|_{-} + {l_A \over 2} G_\mu^\nu (h^A)  =
    {\kappa^2 \over 2} T_\mu^{A\nu} \ .
    \label{JA: atA}
\end{equation}
The junction condition at C-brane is
\begin{eqnarray}
    & & {\chi_\mu^{\nu}|_{+} \over e^{4d_B /l_A}} - {\chi_\mu^{\nu}|_{-} \over
    e^{-4d_A/l_A}}
    +  \left( \alpha - 1 \right) {l_A \over 2} e^{2d_A/l_A} G_\mu^\nu(h^A) \nonumber\\
    && + \left( \alpha - 1 \right) e^{2d_A/l_A}  \left[ \left( \nabla_\mu \nabla^\nu d_A
        - \delta_\mu^\nu \nabla_\sigma \nabla^\sigma d_A \right)  + {1 \over  l_A} \left( \nabla_\mu d_A \nabla^\nu d_A
        + {1\over 2} \delta_\mu^\nu \nabla_\sigma d_A \nabla^\sigma d_A
        \right) \right] \nonumber\\
        && = {\kappa^2 } e^{2d_A/l_A} T_\mu^{C\nu} \ ,
        \label{JA: atC}
\end{eqnarray}
and the junction condition at B-brane yields
\begin{eqnarray}
    & & \chi_\mu^{\nu}|_{+} + {\alpha l_A \over 2} e^{2d_A/l_A+2d_B/l_B} G_\mu^\nu(h^A) \nonumber\\
    & & + \alpha e^{2d_A/l_A+2d_B/l_B} \left[ \left( \nabla_\mu \nabla^\nu d_A
        - \delta_\mu^\nu \nabla_\sigma \nabla^\sigma d_A \right)  + {1 \over  l_A} \left( \nabla_\mu d_A \nabla^\nu d_A
        + {1\over 2} \delta_\mu^\nu \nabla_\sigma d_A \nabla^\sigma d_A
        \right) \right] \nonumber\\
    & & +e^{2d_A/l_A+2d_B/l_B} \left[ \left( \nabla_\mu \nabla^\nu d_B
        - \delta_\mu^\nu \nabla_\sigma \nabla^\sigma d_B \right)  + {1 \over  \alpha l_A} \left( \nabla_\mu d_B \nabla^\nu d_B
        + {1\over 2} \delta_\mu^\nu \nabla_\sigma d_B \nabla^\sigma d_B
        \right) \right] \nonumber \\
    & & + {2 \over l_A} e^{2d_A/l_A+2d_B/l_B}  \left( \nabla_\mu d_A \nabla^\nu d_B
        + {1\over 2} \delta_\mu^\nu \nabla_\sigma d_A \nabla^\sigma d_B
        \right)  \nonumber \\
    & & = - {\kappa^2  \over 2 } e^{2d_A/l_A+2d_B/l_B} T_\mu^{B\nu} \ ,
        \label{JA: atB}
\end{eqnarray}
where $\nabla$ is a derivative covariant with respect to the
induced metric on the A-brane. Eliminating $\chi_\mu^{\nu}|_{-} $
and $\chi_\mu^{\nu}|_{+} $, the equations of motion on the A-brane
is given by
\begin{eqnarray}
    G_\mu^\nu(h^A) &=& {2\kappa^2 \over l_A \left( \Phi_A +
    \alpha (1-\Phi_A) \Psi_A \right)} \left[ {1\over 2}T_\mu^{A\nu} +
    (1-\Phi_A) \left( T_\mu^{C\nu} + {1\over 2} (1-\Psi_A) T_\mu^{B\nu}
    \right) \right] \nonumber\\
    && + {l_A \over 2 \left( \Phi_A +
    \alpha (1-\Phi_A) \Psi_A \right)} \bigg[\alpha (1-\Phi_A) P_\mu^\nu (\Psi_A) + \left( 1 - \alpha \Psi_A \right) P_\mu^\nu (\Phi_A)
    \nonumber\\
    && +\alpha P_\mu^\nu (\Phi_A, \Psi_A) \bigg] \ ,
    \label{eom: Abrane}
\end{eqnarray}
where
\begin{eqnarray}
    P_\mu^\nu(\Phi_A)&=& \left( \nabla_\mu \nabla^\nu \Phi_A
    - \delta_\mu^\nu \nabla_\sigma \nabla^\sigma \Phi_A \right)  \nonumber \\
    && + {3 \over  2(1-\Phi_A)} \bigg( \nabla_\mu \Phi_A\nabla^\nu \Phi_A
    - {1\over 2} \delta_\mu^\nu \nabla_\sigma \Phi_A \nabla^\sigma \Phi_A \bigg)  \ ,
     \label{defP:PhiA} \\
    P_\mu^\nu(\Psi_A)&=& \left( \nabla_\mu \nabla^\nu \Psi_A
    - \delta_\mu^\nu \nabla_\sigma \nabla^\sigma \Psi_A \right)  \nonumber \\
    && + {3 \over  2(1-\Psi_A)} \bigg( \nabla_\mu \Psi_A\nabla^\nu \Psi_A
    - {1\over 2} \delta_\mu^\nu \nabla_\sigma \Psi_A \nabla^\sigma \Psi_A
    \bigg) \ ,
    \label{defP:PsiA} \\
    P_\mu^\nu(\Phi_A,\Psi_A)&=& \nabla_\mu \Phi_A\nabla^\nu \Psi_A
    + {1\over 2} \delta_\mu^\nu \nabla_\sigma \Phi_A \nabla^\sigma
    \Psi_A\ .
    \label{defP:PhiAPsiA}
\end{eqnarray}
Here, we have defined two scalar fields as follows
\begin{equation}
    \Phi_A \equiv 1 - e^{- 2d_A/l_A} \ , \qquad  \Psi_A \equiv 1 -
    e^{-2d_B/l_B} \ .
\end{equation}
Substituting (\ref{eom: Abrane}) into (\ref{JA: atA}) and
(\ref{JA: atC}), respectively, we find
\begin{eqnarray}
    \chi_\mu^{\nu}|_{-} &=&  -  {\kappa^2 (1-\Phi_A) (1-\alpha \Psi_A)\over {\left( \Phi_A +
    \alpha (1-\Phi_A) \Psi_A \right)}} \left[ {1\over 2}T_\mu^{A\nu} +
    {1 \over (1-\alpha \Psi_A)} \left( T_\mu^{C\nu} + {1\over 2} (1-\Psi_A) T_\mu^{B\nu}
    \right)\right]\nonumber\\
    && - { l_A \over 2 {\left( \Phi_A +
    \alpha (1-\Phi_A) \Psi_A \right)}} \Bigg[ \alpha \left( 1 -  \Phi_A \right) P_\mu^\nu(\Psi_A)\nonumber\\
    && +  \left( 1 - \alpha \Psi_A \right)
    P_\mu^\nu(\Phi_A) +  \alpha  P_\mu^\nu(\Phi_A,\Psi_A) \Bigg] \ ,
    \label{chi:negA} \\
    \chi_\mu^{\nu}|_{+} &=& - {\alpha \kappa^2 \over \left( \Phi_A + \alpha(1-\Phi_A)\Psi_A
    \right)(1-\Psi_A)(1-\Phi_A)} \Bigg[ {1\over 2}T_\mu^{A\nu} +
     (1-\Phi_A) T_\mu^{C\nu} \nonumber\\
    &&  + {1 \over 2} { \alpha(1-\Phi_A)+ \Phi_A \over
    \alpha}T_\mu^{B\nu} \Bigg] \nonumber\\
    && - {\alpha l_A \over 2\left( \Phi_A + \alpha(1-\Phi_A)\Psi_A
    \right)(1-\Psi_A)(1-\Phi_A)} \Bigg[ {1\over (1-\Phi_A)} P_\mu^\nu(\Phi_A) \nonumber\\
    && + \Bigg({\alpha(1-\Phi_A) + \Phi_A  \over (1-\Psi_A)} \Bigg) \Bigg( P_\mu^\nu(\Psi_A) + {1\over (1-\Phi_A)}P_\mu^\nu(\Phi_A, \Psi_A)\Bigg)
    \Bigg] \ .
    \label{Chi:postA}
\end{eqnarray}

Using the conditions $\chi_\mu^{\mu}|_{-} = 0$ and
$\chi_\mu^{\mu}|_{+} = 0$ we find the effective equations of
motions for the scalar fields $\Phi_A$ and $\Psi_A$ as follows
\begin{eqnarray}
    & & P_\mu^\mu(\Phi_A) = -{2\kappa^2(1-\Phi_A) \over l_A(1-\alpha)}
    \left[{1\over 2} (1-\alpha) T^A + T^C \right] \ ,
    \label{eom:phiA}\\
    & & P_\mu^\mu(\Psi_A) + {1\over (1-\Phi_A)} P_\mu^\mu(\Phi_A,
    \Psi_A) = {2\kappa^2(1-\Psi_A) \over l_A(1-\alpha)} \left[ T^C + {1 \over 2} \left( {\alpha-1\over \alpha} \right)T^B
    \right] \ .
    \label{eom:psiA}
\end{eqnarray}

Second, we now derive the equations of motion on the B-brane.
Using the relation of the metric between C-brane and B-brane the
junction conditions at the A-brane is
\begin{eqnarray}
    & & \chi_\mu^{\nu}|_{-} + {l_A \over 2} e^{-2d_A/l_A-2d_B/l_B} G_\mu^\nu(h^B) \nonumber\\
    & & - {l_A \over l_B} e^{-2d_A/l_A-2d_B/l_B} \left[ \left( \hat{\nabla}_\mu \hat{\nabla}^\nu d_B
     - \delta_\mu^\nu \hat{\nabla}_\sigma \hat{\nabla}^\sigma d_B \right)  - {1 \over  l_B} \left( \hat{\nabla}_\mu d_B \hat{\nabla}^\nu d_B
        + {1\over 2} \delta_\mu^\nu \hat{\nabla}_\sigma d_B \hat{\nabla}^\sigma d_B
        \right) \right] \nonumber\\
    & & +e^{-2d_A/l_A-2d_B/l_B} \left[ \left( \hat{\nabla}_\mu \hat{\nabla}^\nu d_A
        - \delta_\mu^\nu\hat{\nabla}_\sigma \hat{\nabla}^\sigma d_A \right)  + {1 \over  l_A} \left( \hat{\nabla}_\mu d_A \hat{\nabla}^\nu d_A
        + {1\over 2} \delta_\mu^\nu \hat{\nabla}_\sigma d_A \hat{\nabla}^\sigma d_A
        \right) \right] \nonumber \\
    & & + {2 \over l_A} e^{-2d_A/l_A-2d_B/l_B}  \left( \hat{\nabla}_\mu d_A \hat{\nabla}^\nu d_B
        + {1\over 2} \delta_\mu^\nu \hat{\nabla}_\sigma d_A \hat{\nabla}^\sigma d_B
        \right)  \nonumber \\
    & & = {\kappa^2  \over 2 } e^{-2d_A/l_A-2d_B/l_B} T_\mu^{A\nu} \ ,
        \label{JB: atA}
\end{eqnarray}
and the junction condition at the C-brane is given by
\begin{eqnarray}
    & & {\chi_\mu^{\nu}|_{+} \over e^{4d_B /l_A}} - {\chi_\mu^{\nu}|_{-} \over
    e^{-4d_A/l_A}}
    + e^{-2d_B/l_B} \left( {l_B \over 2} - {l_A \over 2} \right)  G_\mu^\nu
    (h^B) \nonumber\\
    && - {2  e^{-2d_B/l_B} \over  l_B} \left( {l_B \over 2} - {l_A \over 2} \right) \left[ \left( \hat{\nabla}_\mu \hat{\nabla}^\nu d_B
        - \delta_\mu^\nu \hat{\nabla}_\sigma \hat{\nabla}^\sigma d_B \right)  - {1 \over  l_B} \left( \hat{\nabla}_\mu d_B \hat{\nabla}^\nu
        d_B
        + {1\over 2} \delta_\mu^\nu \hat{\nabla}_\sigma d_B \hat{\nabla}^\sigma
        d_B
        \right) \right] \nonumber\\
        && = {\kappa^2 } e^{-2d_B/l_B} T_\mu^{C\nu} \ ,
        \label{JB: atC}
\end{eqnarray}
where $\hat{\nabla}$ is the covariant derivative with respect to
B-brane, while the junction condition at B-brane is
\begin{equation}
    \chi_\mu^{\mu}|_{+} + {\alpha l_A \over 2} G_\mu^\nu (h^B)  = -  {\kappa^2 \over 2} T_\mu^{B\nu} \ .
    \label{JB: atB}
\end{equation}

Eliminating $\chi_\mu^\nu|_{-}$ and $\chi_\mu^\nu|_{+}$ from
equation (\ref{JB: atA})-(\ref{JB: atB}), the equations of motion
on the B-brane is given by
\begin{eqnarray}
    G_\mu^\nu(h^B) &=& {2 \kappa^2 \over l_A {\left( \Phi_B (\Psi_B +1)+
    \alpha  \Psi_B \right)}} \Bigg[ {1\over 2}T_\mu^{B\nu} +
    (\Psi_B+1)  T_\mu^{C\nu} \nonumber\\
    && + {1\over 2} (\Psi_B+1)(\Phi_B+1) T_\mu^{A\nu}
     \Bigg] \nonumber\\
    && + {1 \over  {\left( \Phi_B (\Psi_B +1)+
    \alpha  \Psi_B \right)}} \Bigg[ (\Psi_B + 1) P_\mu^\nu
    (\Phi_B)
    + (\Phi_B + \alpha) P_\mu^\nu (\Psi_B) \nonumber\\
    && - P_\mu^\nu (\Phi_B,\Psi_B) \Bigg] \ ,
    \label{eom: Bbrane}
\end{eqnarray}
where
\begin{eqnarray}
    P_\mu^\nu(\Phi_B)&=& \left( \hat{\nabla}_\mu \hat{\nabla}^\nu \Phi_B
    - \delta_\mu^\nu \hat{\nabla}_\sigma \hat{\nabla}^\sigma \Phi_B \right)  \nonumber \\
    && - {3 \over  2(\Phi_B+1)} \bigg( \hat{\nabla}_\mu \Phi_B\hat{\nabla}^\nu \Phi_B
    - {1\over 2} \delta_\mu^\nu \hat{\nabla}_\sigma \Phi_B \hat{\nabla}^\sigma \Phi_B \bigg)  \ ,
     \label{defP:PhiB} \\
    P_\mu^\nu(\Psi_B)&=& \left( \hat{\nabla}_\mu \nabla^\nu \Psi_B
    - \delta_\mu^\nu \hat{\nabla}_\sigma \hat{\nabla}^\sigma \Psi_B \right)  \nonumber \\
    && - {3 \over  2(\Psi_B+1)} \bigg( \hat{\nabla}_\mu \Psi_B \hat{\nabla}^\nu \Psi_B
    - {1\over 2} \delta_\mu^\nu \hat{\nabla}_\sigma \Psi_B \hat{\nabla}^\sigma \Psi_B \bigg) \ ,
    \label{defP:PsiB} \\
    P_\mu^\nu(\Phi_B,\Psi_B)&=& \hat{\nabla}_\mu \Phi_B\hat{\nabla}^\nu\Psi_B
    + {1\over 2} \delta_\mu^\nu \hat{\nabla}_\sigma \Phi_B \hat{\nabla}^\sigma
    \Psi_B\ .
    \label{defP:PhiAPsiB}
\end{eqnarray}
The two scalar fields are defined as follows
\begin{equation}
    \Phi_B \equiv  e^{ 2d_A/l_A} - 1 \ , \qquad  \Psi_B \equiv  e^{2d_B/l_B} - 1 \ .
\end{equation}
The solutions for $\chi_\mu^\nu|_{-}$ and $\chi_\mu^\nu|_{+}$ can
be obtained by substituting equation (\ref{eom: Bbrane}) into
(\ref{JB: atA}) and (\ref{JB: atB}), respectively. Then, we get
\begin{eqnarray}
    {\chi_\mu^{\nu}|_{-}} &=& -{\kappa^2 \over {\left( \Phi_B (\Psi_B +1)+
    \alpha  \Psi_B \right)} (\Phi_B+1)(\Psi_B+1)} \Bigg[{1\over2}
    T_\mu^{B\nu} + (\Psi_B + 1) T_\mu^{C\nu} \nonumber \\
    && + {1\over2}(\Psi_B+1) \left(1-{\alpha \Psi_B \over (\Psi_B+1)}
    \right) T_\mu^{A\nu} \Bigg] \nonumber\\
    && - {l_A \over 2 \left( \Phi_B (\Psi_B +1)+
    \alpha  \Psi_B \right)(\Phi_B+1)^2(\Psi_B+1)^2} \Bigg[ \alpha (\Phi_B+1) P_\mu^\nu(\Psi_B) \nonumber\\
    &&+ \left( (\Psi_B+1)- \alpha \Psi_B \right) \left( (\Psi_B+1)P_\mu^\nu(\Phi_B)
     -P_\mu^\nu(\Phi_B,\Psi_B) \right) \Bigg] \ ,
     \label{chi:negB} \\
    {\chi_\mu^{\nu}|_{+}} &=& - {\kappa^2 (\Psi_B+1)(\alpha+\Phi_B) \over \left( \Phi_B (\Psi_B +1)+
    \alpha  \Psi_B \right)} \Bigg[{1\over2} T_\mu^{B\nu} + \Bigg( {\alpha \over \alpha + \Phi_B} \Bigg)
    T_\mu^{C\nu} \nonumber\\
    &&  + {1\over2} \Bigg( {\alpha \over \alpha + \Phi_B}\Bigg) (\Phi_B+1)T_\mu^{A\nu}
    \Bigg]\nonumber\\
    && - { \alpha l_A \over 2 \left( \Phi_B (\Psi_B +1)+
    \alpha  \Psi_B \right)} \Bigg[ (\Psi_B+1) P_\mu^\nu(\Phi_B) + (\Phi_B + \alpha )P_\mu^\nu(\Psi_B) \nonumber\\
    &&- P_\mu^\nu(\Phi_B,\Psi_B) \Bigg] \ .
    \label{chi:posB}
\end{eqnarray}

Finally, the equations of motion for the scalar fields are given
by
\begin{eqnarray}
    & & P_\mu^\mu(\Phi_B) - {1\over (\Psi_B +1)} P_\mu^\mu(\Phi_B,
    \Psi_B)= -{2\kappa^2(\Phi_B+1) \over l_A}
    \left[{1\over 2} T^A + {1\over (1-\alpha)}T^C \right] \ ,
    \label{eom:phiB} \\
    & & P_\mu^\mu(\Psi_B)  = - {2\kappa^2(\Psi_B+1) \over l_A(\alpha-1)} \left[ {1\over 2}\left({\alpha-1 \over \alpha} \right)T^B
    +  T^C  \right] \ .
    \label{eom:psiB}
\end{eqnarray}

In the derivation of equations of motion above we first to know
the dynamics on one brane. Then we know the gravity on the other
branes. Since the dynamics on each branes are not independent, the
transformation rules for the scalar fields are given by
\begin{eqnarray}
    \Phi_C = \Phi_B \ = {\Phi_A \over 1-\Phi_A} \ , \qquad  \Psi_C = \Psi_A = {\Psi_B \over 1 + \Psi_B} \ .
    \label{rule:field}
\end{eqnarray}

In the following subsection, for the realization at the first
order expansion, we study the cosmological consequence of the
radion dynamics. We need to derive the Friedmann equation on each
branes.

\subsection{The effective Friedmann equation on the branes}

The metric induced on the brane is the Friedmann-Robertson-Walker
(FRW) metric,
\begin{equation}
    ds^2 = -dt^2 + a^2_i(t) \gamma_{mn} dx^m dx^n \ , \qquad i = A, B, C
    \label{FRWmetric}
\end{equation}
where the time and space components of the Einstein equations are
given by
\begin{eqnarray}
    G^0_0 &=& -3 \left[ {{\dot{a}^2_i + k} \over a^2_i} \right] \ ,
    \label{time-comp} \\
    G_n^m &=&  -  \left[ {{2 a_i \ddot{a}_i + \dot{a}^2_i + k} \over a^2_i} \right] \delta_n^m  \
    ,
    \label{spatial-comp}
\end{eqnarray}
and $k$ is the spatial curvature, $k=0, \pm 1$.

Assuming that the matter of the energy-momentum tensor to be
$T^{i^\mu}_\nu = -\rho^i \delta^\mu_\nu $, $(i=A, B, C)$, then the
field equations on the middle brane can be written as
\begin{eqnarray}
    G^0_0(h^C) &=& - {{2\kappa^2 \rho^C} \over {l_A (\Phi_C + \alpha    \Psi_C)}}
    \left[ 1 + {(1+ \Phi_C) \over 2\alpha_A} + {(1 - \Psi_C) \over 2\alpha_B}  \right] \nonumber \\
    && + {3 \over  (\Phi_C + \alpha  \Psi_C)} \left[ H (\dot{\Phi}_C + \alpha  \dot{\Psi}_C) + {1\over 4}\left( {{\dot{\Phi}_C}^2 \over
    (1+\Phi_C)} -{\alpha{\dot{\Psi}}_C^2 \over (1-\Psi_C)} \right)\right] \ ,
    \label{time-eom} \\
    G^m_m &=& - {{2\kappa^2 \rho^C} \over {l_A (\Phi_C + \alpha    \Psi_C)}}
    \left[ 1 + {(1+ \Phi_C) \over 2\alpha_A} + {(1 - \Psi_C) \over 2\alpha_B}  \right] \nonumber \\
    && + {1 \over  (\Phi_C + \alpha  \Psi_C)} \Bigg[ 2H (\dot{\Phi}_C + \alpha  \dot{\Psi}_C) + (\ddot{\Phi}_C + \alpha
    \ddot{\Psi}_C) \nonumber \\
     &&- {3\over 4}\Bigg( {{\dot{\Phi}}_C^2 \over (1+\Phi_C)} - {\alpha{\dot{\Psi}}_C^2 \over (1-\Psi_C)} \Bigg)\Bigg] \
     ,
    \label{spatial-eom}
\end{eqnarray}
where the equations of motion for the radion fields are
\begin{eqnarray}
    {\ddot{\Phi}_C} + 3H \dot{\Phi}_C &=& - {{8 \kappa^2 \rho^C (1+\Phi_C)} \over {3 l_A }}
    \left[ {1 \over (\alpha - 1) } - {1\over 2\alpha_A} \right] + {{\dot{\Phi}}_C^2 \over 2(1+\Phi_C)} \ ,
    \label{eom-Phi_C} \\
    {\ddot{\Psi}_C} + 3H \dot{\Psi}_C &=& {{8 \kappa^2 \rho^C (1-\Psi_C)} \over {3 l_A }}
    \left[ {1 \over (\alpha -1)} + {1 \over 2\alpha_B} \right] -  {\dot{\Psi}_C^2 \over 2(1-\Psi_C)}  \ .
     \label{eom-Psi_C}
\end{eqnarray}
Here, we have defined two dimensionaless parameters
\begin{equation}
    \alpha_A = {\rho^C \over \rho^A}, \qquad \alpha_B = {\rho^C \over
    \rho^B}\ .
\end{equation}
Using the equations (\ref{time-comp}) and (\ref{spatial-comp}) and
eliminating $\ddot{\Phi}$ and $\ddot{\Psi}$ from (\ref{eom-Phi_C})
and (\ref{eom-Psi_C}), respectively, we get
\begin{equation}
     \dot{H} + 2H^2 + {k\over a^2_C} = {{4 \kappa^2   \over {3 l_A (\alpha -1)}} \rho^C} \ . \label{FRW-eq}
\end{equation}
Integrating this equation, we obtain the Friedmann equation with
dark radiation,
\begin{equation}
     H^2 + {k\over a^2_C} = {{2 \kappa^2   \over {3 l_A (\alpha -1)}} \rho^C} + {C_C\over a^4_C} \ , \label{FRW-sol}
\end{equation}
where $C_C$ is an integration constant.

We derive the Friedmann equation on the A-brane where the FRW
metric is given by (\ref{FRWmetric}). From the equation (\ref{eom:
Abrane}) the time component of the Einstein equation is
\begin{eqnarray}
    G_0^0 (h^A) &=& - {\kappa^2 \rho^C \over l\theta} \left[ {1\over 2 \alpha_A} +
    (1-\Phi_A) \left( 1 + {1\over 2 \alpha_B} (1-\Psi_A) \right) \right] \nonumber\\
    && + {l_A \over 2 l \theta} \bigg[\alpha (1-\Phi_A) P_0^0 (\Psi_A) + \left( 1 - \alpha \Psi_A \right) P_0^0 (\Phi_A)
     +\alpha P_0^0 (\Phi_A, \Psi_A) \bigg] \ ,
    \label{eom: Abrane00}
\end{eqnarray}
where $l\theta$ is defined as follows
\begin{equation}
    l \theta \equiv {l_A \over 2} \left( \Phi_A + \alpha (1-\Phi_A) \Psi_A
    \right) \ .
\end{equation}
And the time component of the kinetic terms are given by
\begin{eqnarray}
    P_0^0 (\Psi_A) &=& 3H \dot{\Psi}_A - {3\over 4(1-\Psi_A)} \dot{\Psi}_A^2 \ ,
    \label{psiA:00}\\
    P_0^0 (\Phi_A) &=& 3H \dot{\Phi}_A - {3\over 4(1-\Phi_A)} \dot{\Phi}_A^2 \  ,
    \label{phiA:00} \\
    P_0^0 (\Phi_A, \Psi_A) &=& - {3\over 2} \dot{\Phi}_A \dot{\Psi}_A \  .
    \label{phiApsiA:00}
\end{eqnarray}

The space component of the Einstein equation (\ref{eom: Abrane})
is given by
\begin{eqnarray}
    G_m^m (h^A) &=& - {\kappa^2 \rho^C \over l\theta} \left[ {1\over 2 \alpha_A} +
    (1-\Phi_A) \left( 1 + {1\over 2 \alpha_B} (1-\Psi_A) \right) \right] \nonumber\\
    && + {l_A \over 2 l \theta} \bigg[\alpha (1-\Phi_A) P_m^m (\Psi_A) + \left( 1 - \alpha \Psi_A \right) P_m^m (\Phi_A)
     +\alpha P_m^m (\Phi_A, \Psi_A) \bigg] \ ,
    \label{eom: Abranemm}
\end{eqnarray}
where
\begin{eqnarray}
    P_m^m (\Psi_A) &=& \ddot{\Psi}_A + 2H \dot{\Psi}_A + {3\over 4(1-\Psi_A)}\dot{\Psi}_A^2 \ ,
    \label{psiA:mm}\\
    P_m^m (\Phi_A) &=& \ddot{\Phi}_A + 2H \dot{\Phi}_A + {3\over 4(1-\Phi_A)}\dot{\Phi}_A^2 \  ,
    \label{phiA:mm} \\
    P_m^m (\Phi_A, \Psi_A) &=& - {1\over 2} \dot{\Phi}_A \dot{\Psi}_A \  .
    \label{phiApsiA:mm}
\end{eqnarray}
Inserting equations (\ref{psiA:00})-(\ref{phiApsiA:00}) into
(\ref{eom: Abrane00}) and (\ref{psiA:mm})-(\ref{phiApsiA:00}) into
(\ref{eom: Abranemm}), respectively, we obtain
\begin{eqnarray}
     -3 \left( H^2 + {k\over a^2_A} \right) &= & - {\kappa^2 \rho^C \over
    l\theta } \left[ {1\over 2 \alpha_A} +
    (1-\Phi_A) \left( 1 + {1\over 2 \alpha_B} (1-\Psi_A) \right) \right] \nonumber\\
    && + {l_A \over 2 l \theta} \bigg[\alpha (1-\Phi_A) \Bigg(3H \dot{\Psi}_A - {3\over 4(1-\Psi_A)}
    \dot{\Psi}_A^2\Bigg) \nonumber\\
    && + (1-\alpha\Psi_A) \Bigg( 3H \dot{\Phi}_A - {3\over 4(1-\Phi_A)} \dot{\Phi}_A^2
    \Bigg) - {3\over 2} \dot{\Phi}_A \dot{\Psi}_A \Bigg] \ ,
    \label{EinstA:00}
\end{eqnarray}
and
\begin{eqnarray}
    && -2\left( \dot{H} - {k\over a^2_A} \right)-3 \left( H^2 + {k\over a^2_A} \right) = - {\kappa^2 \rho^C \over
    l\theta } \left[ {1\over 2 \alpha_A} +
    (1-\Phi_A) \left( 1 + {1\over 2 \alpha_B} (1-\Psi_A) \right) \right] \nonumber\\
    && \qquad \qquad \qquad +{l_A \over 2 l \theta} \bigg[\alpha (1-\Phi_A) \Bigg(\ddot{\Psi}_A + 2H \dot{\Psi}_A + {3\over
    4(1-\Psi_A)}\dot{\Psi}_A^2 \Bigg) \nonumber\\
    && \qquad \qquad \qquad + (1-\alpha \Psi_A) \Bigg(   \ddot{\Phi}_A + 2H \dot{\Phi}_A + {3\over 4(1-\Phi_A)}\dot{\Phi}_A^2  \Bigg) - {1\over 2} \dot{\Phi}_A \dot{\Psi}_A
    \Bigg]\ ,
    \label{EinstA:mm}
\end{eqnarray}
where the equations of motion for the radion fields $\Phi_A$ and
$\Psi_A$ are obtained from the equations (\ref{eom:phiA}) and
(\ref{eom:psiA}), respectively
\begin{eqnarray}
    \ddot{\Phi}_A &=& {8\kappa^2 \rho^C (1-\Phi_A) \over 3l_A(1-\alpha)}
    \left[{(1-\alpha) \over 2 \alpha_A} + 1 \right] - 3H\dot{\Phi}_A - {1\over 2(1-\Phi_A)} \dot{\Phi}_A^2  \ ,
    \label{eom:phiAexplisit}\\
    \ddot{\Psi}_A &=& - {8\kappa^2 \rho^C (1-\Psi_A) \over 3l_A(1-\alpha)}
    \left[1 - {(1-\alpha) \over 2 \alpha \alpha_B} \right] - 3H\dot{\Psi}_A - {1\over 2(1-\Psi_A)}
    \dot{\Psi}_A^2\nonumber\\
    &&+ {1\over 2(1-\Phi_A)} \dot{\Phi}_A \dot{\Psi}_A \ .
    \label{eom:psiAexplixit}
\end{eqnarray}
Substituting (\ref{eom:phiAexplisit}) and (\ref{eom:psiAexplixit})
into (\ref{EinstA:mm}), respectively, we obtain
\begin{equation}
    \dot{H} +2H^2 + {k\over a^2_A} = {2\kappa^2 \rho^A \over
    3l_A} \ .
    \label{eom:FriedmannA}
\end{equation}
Then, we obtain the Friedmann equation with dark radiation by
integrating (\ref{eom:FriedmannA}) on the A-brane as follows
\begin{equation}
    H^2 + {k\over a^2_A} = {\kappa^2 \rho^A \over 3l_A} + {C_A \over a^4_A}  \ ,
    \label{Friedmann:EqA}
\end{equation}
where $C_A$ is an integration constant.

Finally, To obtain the Friedmann equation on the B-brane we use
the same procedure. The Einstein equations on B-brane are given as
follows
\begin{eqnarray}
     -3 \left( H^2 + {k\over a^2_B} \right) &= & - {\kappa^2 \rho^C \over
    \widetilde{l\theta} } \left[ {1\over 2 \alpha_B} +
    (1+\Psi_B) \left( 1 + {1\over 2 \alpha_A} (1+\Phi_B) \right) \right] \nonumber\\
    && + {l_A \over 2 \widetilde{l\theta}} \bigg[ (1+\Psi_B) \Bigg(3H \dot{\Phi}_B + {3\over 4(1+\Phi_B)}
    \dot{\Phi}_B^2\Bigg) \nonumber\\
    && + (\alpha+\Phi_B) \Bigg( 3H \dot{\Psi}_B + {3\over 4(1+\Psi_B)} \dot{\Psi}_B^2
    \Bigg) + {3\over 2} \dot{\Phi}_A \dot{\Psi}_A \Bigg] \ ,
    \label{EinstB:00}
\end{eqnarray}
and
\begin{eqnarray}
    && -2\left( \dot{H} - {k\over a^2_B} \right)-3 \left( H^2 + {k\over a^2_B} \right) = - {\kappa^2 \rho^C \over
    \widetilde{l\theta} } \left[ {1\over 2 \alpha_B} +
    (1+\Psi_B) \left( 1 + {1\over 2 \alpha_A} (1+\Phi_B) \right) \right] \nonumber\\
    && \qquad \qquad \qquad +{l_A \over 2 \widetilde{l\theta}} \bigg[(1+\Psi_B) \Bigg(\ddot{\Phi}_B + 2H \dot{\Phi}_B - {3\over
    4(1+\Phi_B)}\dot{\Phi}_B^2 \Bigg) \nonumber\\
    && \qquad \qquad \qquad + (\alpha+\Phi_B) \Bigg(   \ddot{\Psi}_B + 2H \dot{\Psi}_B - {3\over 4(1+\Psi_B)}\dot{\Psi}_B^2  \Bigg) + {1\over 2} \dot{\Phi}_A \dot{\Psi}_A
    \Bigg]\ ,
    \label{EinstB:mm}
\end{eqnarray}
where $\widetilde{l\theta}$ is defined as follows
\begin{equation}
    \widetilde{l\theta} \equiv {l_A \over 2} \left( \Phi_B(\Psi_B + 1) + \alpha \Psi_B \right) \ .
\end{equation}
In order to obtain (\ref{EinstB:00}) and (\ref{EinstB:mm}) we have
used the components of the scalar kinetic terms:
\begin{eqnarray}
    P_0^0 (\Psi_B) &=& 3H \dot{\Psi}_B + {3\over 4(1+\Psi_B)} \dot{\Psi}_B^2 \ ,
    \label{psiB:00}\\
    P_0^0 (\Phi_B) &=& 3H \dot{\Phi}_B + {3\over 4(1+\Phi_B)} \dot{\Phi}_B^2 \  ,
    \label{phiB:00} \\
    P_0^0 (\Phi_B, \Psi_B) &=& - {3\over 2} \dot{\Phi}_B \dot{\Psi}_B ,
    \label{phiBpsiB:00} \\
    P_m^m (\Psi_B) &=& \ddot{\Psi}_B + 2H \dot{\Psi}_B - {3\over 4(1+\Psi_B)}\dot{\Psi}_B^2 \ ,
    \label{psiB:mm}\\
    P_m^m (\Phi_B) &=& \ddot{\Phi}_B + 2H \dot{\Phi}_B - {3\over 4(1+\Phi_B)}\dot{\Phi}_B^2 \  ,
    \label{phiB:mm} \\
    P_m^m (\Phi_B, \Psi_B) &=& - {1\over 2} \dot{\Phi}_B \dot{\Psi}_B \  .
    \label{phiBpsiB:mm}
\end{eqnarray}

The equations of motion for the scalar fields $\Phi_B$ and
$\Psi_B$ are used to eliminate the second derivative
$\ddot{\Phi}_B$ and $\ddot{\Psi}_B$ in the equation
(\ref{EinstB:mm}), where it is given by
\begin{eqnarray}
    \ddot{\Psi}_B &=& {8\kappa^2 \rho^C (1+\Psi_B) \over 3l_A(\alpha-1)}
    \left[{(\alpha-1) \over 2 \alpha \alpha_B} + 1 \right] - 3H\dot{\Psi}_B + {1\over 2(1+\Psi_B)} \dot{\Psi}_B^2  \ ,
    \label{eom:phiBexplisit}\\
    \ddot{\Phi}_B &=&  {8\kappa^2 \rho^C (1+\Phi_B) \over 3l_A}
    \left[{1\over 2 \alpha_A} + {1\over (1-\alpha)} \right] - 3H\dot{\Phi}_B + {1\over 2(1+\Phi_B)}
    \dot{\Phi}_B^2\nonumber\\
    &&- {1\over 2(1+\Psi_B)} \dot{\Phi}_B \dot{\Psi}_B \ .
    \label{eom:psiBexplixit}
\end{eqnarray}
Solving the equation (\ref{EinstB:mm}) we get the Friedmann
equation on the B-brane by integrating the equation below
\begin{equation}
    \dot{H} +2H^2 + {k\over a^2_B} = -{2\kappa^2 \rho^B \over
    3l_B} \ ,
    \label{eom:FriedmannB}
\end{equation}
to find
\begin{equation}
    H^2 + {k\over a^2_B} = - {\kappa^2 \rho^B \over
    3l_B } + {C_B \over a^4_B} \ ,
    \label{eom:FriedmannA}
\end{equation}
where $C_B$ is an integration constant.

\section{The scalar-tensor gravity}

In the previous section we have derived the effective equations of
motion in a three brane system. Now we show how we can write the
scalar-tensor gravity using the effective equations of motion on
this system. We use the solutions on the middle brane to obtain a
scalar-tensor gravity with two independent scalar fields. In the
following we omit subscript $C$ of the equations that related to
the middle brane. From the equation (\ref{effEq:C}) we see that a
term ${l_A \over 2} \left( \Phi + \alpha \Psi \right)$ can be
defined as a first dimensionless scalar field,
\begin{equation}
    l \phi \equiv {l_A \over 2} \left( \Phi + \alpha \Psi \right) \ ,
    \label{def:new scalar1}
\end{equation}
where $l$ is an arbitrary unit of length. Because the scalar
fields $\Phi$ and $\Psi$ correspond to the proper distance, the
definition of the scalar field (\ref{def:new scalar1}) associated
with overall distance of the middle brane. Then, the second scalar
field can also defined as a function of both scalar fields, we
define
\begin{equation}
    \varphi \equiv \varphi (\xi(\Phi,\Psi)) \ .
    \label{def:new scalar2}
\end{equation}

We intend to write the effective equations of motion on the middle
brane (\ref{effEq:C}) as follows
\begin{eqnarray}
    G_\mu^\nu &=&
    { \kappa^2 \over {l \phi} } \left(  {{1 + \Phi} \over 2} T_\mu^{A\nu} + {{1 - \Psi} \over 2}
    T_\mu^{B\nu} + T_\mu^{C\nu} \right) + {1\over \phi} \left( \mathcal{D}_\mu \mathcal{D}^\nu \phi -
    \delta_\mu^{\nu} \Box \phi \right) \nonumber \\
    && + {\omega(\phi) \over \phi } \left[ \left( \mathcal{D}_\mu \phi \mathcal{D}^\nu \phi
    - {1\over 2} \delta_\mu^{\nu} (\mathcal{D} \phi)^2\right) - \bar{\omega}(\phi) \left( \mathcal{D}_\mu \varphi \mathcal{D}^\nu \varphi
    - {1\over 2} \delta_\mu^{\nu} (\mathcal{D} \varphi)^2 \right)\right],
    \label{eq-likeBD}
\end{eqnarray}
where $\omega(\phi)$ and $\overline{\omega}(\phi)$ are the
arbitrary functional coupling of $\phi$. The absence of mixing
terms in the equation above yields the following constraints
\begin{equation}
    \alpha \left({l_A \over 2l}\right)^2
    - \bar{\omega}(\phi) \left({d \varphi \over d\xi} \right)^2 \left({\partial\xi \over \partial \Phi}{\partial\xi \over \partial \Psi}\right)=0 \ .
    \label{constraint}
\end{equation}
Applying this constraint into equation (\ref{eq-likeBD}) we have
\begin{eqnarray}
    G_\mu^\nu &=&
    { \kappa^2 \over {l \phi} } \left(  {{1 + \Phi} \over 2} T_\mu^{A\nu} + {{1 - \Psi} \over 2}
    T_\mu^{B\nu} + T_\mu^{C\nu} \right) + {1\over \phi} \left( \mathcal{D}_\mu \mathcal{D}^\nu \phi -
    \delta_\mu^{\nu} \Box \phi \right) \nonumber \\
    && +{\omega(\phi) \over \phi }  \left[ \left({l_A \over 2l} \right)^2 - \overline{\omega}(\phi) \left({d\varphi \over d\xi} {\partial \xi \over \partial \Phi} \right)^2\right]
    \left( \mathcal{D}_\mu \Phi \mathcal{D}^\nu \Phi
    - {1\over 2} \delta_\mu^{\nu} (\mathcal{D} \Phi)^2 \right)
    \nonumber \\
    && + {\omega(\phi) \over \phi } \left[ \left({\alpha l_A \over 2l} \right)^2 - \overline{\omega}(\phi) \left({d\varphi \over d\xi} {\partial \xi \over \partial \Psi} \right)^2\right]
    \left( \mathcal{D}_\mu \Psi \mathcal{D}^\nu \Psi
    - {1\over 2} \delta_\mu^{\nu} (\mathcal{D} \Psi)^2 \right) \ .
    \label{eq-likeBD1}
\end{eqnarray}
On the other hand, by inserting (\ref{def:new scalar1}) into
(\ref{effEq:C}) we obtain
\begin{eqnarray}
    G_\mu^\nu &=&
    { \kappa^2 \over {l \phi} } \left(  {{1 + \Phi} \over 2} T_\mu^{A\nu} + {{1 - \Psi} \over 2}
    T_\mu^{B\nu} + T_\mu^{C\nu} \right) + {1\over \phi} \left( \mathcal{D}_\mu \mathcal{D}^\nu \phi -
    \delta_\mu^{\nu} \Box \phi \right) \nonumber \\
    && - {3 l_A \over {4 l \phi (1+\Phi)} } \left( \mathcal{D}_\mu \Phi \mathcal{D}^\nu \Phi
    - {1\over 2} \delta_\mu^{\nu} (\mathcal{D} \Phi)^2 \right)
    \nonumber \\
    && +{3 \alpha l_A \over {4 l \phi (1 -\Psi)} }\left( \mathcal{D}_\mu \Psi \mathcal{D}^\nu \Psi
    - {1\over 2} \delta_\mu^{\nu} (\mathcal{D} \Psi)^2 \right) \ .
    \label{new: effEq C baru}
\end{eqnarray}
The relation between the coefficients in equation
(\ref{eq-likeBD1}) and (\ref{new: effEq C baru}) is
\begin{equation}
    {3 l_A \over {4 l \phi (1+\Phi)} } ={\omega(\phi) \over \phi
    }\left({l_A \over 2l} \right)^2 \left[ \alpha
    {{\partial\xi/\partial \Phi} \over {\partial\xi/\partial
    \Psi}}-1
    \right] ={\omega(\phi) \over \phi} \left({l_A \over 2l} \right)^2 \left[ \alpha -
    {{\partial\xi/\partial \Psi} \over {\partial\xi/\partial
    \Phi}}  \right] \ .
    \label{relatcof}
\end{equation}
and using the constraint (\ref{constraint}) we get an equation of
the form

\begin{equation}
    \left[ 1 - {(1+\Phi)\over (1-\Psi)}{\partial\xi /\partial \Phi \over \partial\xi /\partial \Psi} \right]
    \left[\alpha - {\partial\xi /\partial \Psi \over \partial\xi /\partial \Phi}
    \right]=0 \ .
    \label{solscalar1}
\end{equation}
It is easy to see that the solution of the equation
(\ref{solscalar1}) are
\begin{eqnarray}
    \partial\xi /\partial \Psi \over \partial\xi /\partial \Phi
    &=&
    \alpha , \\
    \partial\xi /\partial \Psi \over \partial\xi /\partial \Phi &=& (1+\Phi)\over
    (1-\Psi) \ .
\end{eqnarray}
The first solution yields vanishing the coefficient of
(\ref{eq-likeBD1}) and (\ref{new: effEq C baru}). Then we find a
solution,
\begin{equation}
    \xi = \log {(1 + \Phi) \over (1 - \Psi)} \ .
    \label{xi-parsol}
\end{equation}
Substituting this solution into (\ref{constraint}) we obtain a
differential equation for $\varphi$
\begin{equation}
    {d \varphi \over d \xi} = { \sqrt{\alpha} e^{\xi/2} \over (e^\xi - \alpha)} \sqrt{{\left( (1 - \alpha) l_A/2 + l \phi  \right)^2
    } \over {l^2} \overline{\omega}}
     \ .
     \label{descalar}
\end{equation}
In order for $\varphi$ to be only a function of $\xi$, we require
that in equation (\ref{descalar})
\begin{equation}
     \left( {(1 - \alpha) l_A \over 2} + l \phi  \right)^2
     = l^2 \overline{\omega}(\phi)  \ .
     \label{descalar2}
\end{equation}
Then, we obtain the solution of $\varphi$,
\begin{equation}
      e^{\xi} = \alpha \coth^2 \left( {\varphi(\xi)\over 2}\right) \ .
     \label{sol:scalar2}
\end{equation}
Finally, the effective equations of motion on the middle brane can
be written as
\begin{eqnarray}
    G_\mu^\nu &=&
    { \kappa^2 \over {l \phi} } \left[  {{3 l \phi \over 2 l_A \omega(\phi)} \cosh^2 \left( {\varphi(\xi)\over 2}\right)} T_\mu^{A\nu}
    + {{3 l \phi \over 2 l_B \omega(\phi)} \sinh^2 \left( {\varphi(\xi)\over 2}\right)}
    T_\mu^{B\nu} + T_\mu^{C\nu} \right] \nonumber \\
    && + {1\over \phi} \left( \mathcal{D}_\mu \mathcal{D}^\nu \phi -
    \delta_\mu^{\nu} \Box \phi \right)
    + {\omega(\phi) \over \phi } \left( \mathcal{D}_\mu \phi \mathcal{D}^\nu \phi
    - {1\over 2} \delta_\mu^{\nu} (\mathcal{D} \phi)^2 \right) \nonumber \\
    && - {9 \over {4 \omega(\phi)}} \left( \mathcal{D}_\mu \varphi \mathcal{D}^\nu \varphi
    - {1\over 2} \delta_\mu^{\nu} (\mathcal{D} \varphi)^2 \right) \ ,
    \label{final-eom}
\end{eqnarray}
where
\begin{equation}
      \omega(\phi) = - {3\over 2} \left( {l \phi \over {l \phi + (1 - \alpha)l_A/2}}
      \right) \ .
\end{equation}

The effective action for the the middle brane corresponding to the
effective equations of motion (\ref{def:new scalar1}) can be
rewrite as
\begin{eqnarray}
    S &=& {l \over {2 \kappa^2}} \int d^4x \sqrt{-h} \left[\phi R
    - {\omega(\phi)\over \phi} \mathcal{D}_\mu \phi \mathcal{D}^\mu
    \phi + {9\over 4} {\phi \over \omega(\phi)}\mathcal{D}_\mu \varphi \mathcal{D}^\mu
    \varphi \right] \nonumber \\
    &&+ \int d^4x \sqrt{-h} \left[ \mathcal{L}^C + {3l \phi \over {2 l_A \omega(\phi)}} \cosh^2 \left( {\varphi \over
    2} \right)  \mathcal{L}^A + {3l \phi \over {2 l_B \omega(\phi)}} \sinh^2 \left( {\varphi \over
    2} \right) \mathcal{L}^B \right] \ ,
    \label{final action}
\end{eqnarray}
where $\mathcal{L}^A$, $\mathcal{L}^C$ and $\mathcal{L}^B$ are the
Lagrangian correspond to A-brane, C-brane and B-brane,
respectively. This action is the scalar-tensor gravity on the
brane with two scalar fields as a function of the two proper
distance.

\section{Conclusions}

In this paper, we consider three 3-brane systems with with A and
B-brane are placed at the fixed point of the orbifold whereas the
C-brane is put between A- and B-brane.

We use the gradient expansion method to analyze, in the first
order, the effective equations of motion, in particular the radion
Lagrangian. In this case we derived the Friedmann equation with
dark energy radiation by direct elimination of the radion fields
in the Einstein equations. We also derive the scalar-tensor
gravity with depend on the scalar (radion) fields.

We can also generalize this scenario to the multi (more than
three) 3-branes system in the low energy limit. It is also
interesting, in this model, to investigate for higher order
correction.

%\appendix

\acknowledgments

We would like to thank Jiro Soda for a useful comments and
suggestions, and for informing us of Ref.~\cite{LCotta}. We would
also like to thank Kazuya Koyama for informing us of his paper
(Ref.~\cite{KK}). One of us (AR) would like to thank BPPS, Dikti,
Depdiknas, Republic of Indonesia for financial support. He also
wishes to acknowledge all members of Theoretical Physics
Laboratory, Department of Physics, ITB, for warmest hospitality.

\end{document}